\documentclass[manuscript]{emulateapj} 
\usepackage{natbib}

\shorttitle{Big Little Planets} 
\shortauthors{Batygin \& Stevenson}

\begin{document}
 
\title{Mass-Radius Relationships for Very Low Mass Gaseous Planets}  
\author{Konstantin Batygin$^{1}$ \& David J. Stevenson$^2$} 

\affil{$^1$Institute for Theory and Computation, Harvard-Smithsonian Center for Astrophysics, 60 Garden St., Cambridge, MA 02138} 
\affil{$^2$Division of Geological and Planetary Sciences, California Institute of Technology, 1200 E. California Blvd., Pasadena, CA 91125}

\email{kbatygin@cfa.harvard.edu}

\keywords{planets and satellites: interiors, planets and satellites: physical evolution}

\begin{abstract}
Recently, the \textit{Kepler} spacecraft has detected a sizable aggregate of objects, characterized by giant-planet-like radii and modest levels of stellar irradiation. With the exception of a handful of objects, the physical nature, and specifically the average densities, of these bodies remain unknown. Here, we propose that the detected giant planet radii may partially belong to planets somewhat less massive than Uranus and Neptune. Accordingly, in this work, we seek to identify a physically sound upper limit to planetary radii at low masses and moderate equilibrium temperatures. As a guiding example, we analyze the interior structure of the Neptune-mass planet \textit{Kepler}-30d and show that it is acutely deficient in heavy elements, especially compared with its solar system counterparts. Subsequently, we perform numerical simulations of planetary thermal evolution and in agreement with previous studies, show that generally, $10 - 20 M_{\oplus}$, multi-billion year old planets, composed of high density cores and extended H/He envelopes can have radii that firmly reside in the giant planet range. We subject our results to stability criteria based on extreme ultraviolet radiation, as well as Roche-lobe overflow driven mass-loss and construct mass-radius relationships for the considered objects. We conclude by discussing observational avenues that may be used to confirm or repudiate the existence of putative low mass, gas-dominated planets.
\end{abstract}

\section{Introduction}
The ever-growing transit data set collected by the $Kepler$ spacecraft has proven to be instrumental to the advancement of our understanding of the properties of planetary systems. Thanks to the sheer size of the data set ($\sim2500$ planetary candidates as of Quarter 6) and the associated statistical ability to determine the characteristics of typical planetary systems \citep{2010Sci...330..653H, 2011ApJ...742...38Y}, as well as highlight some unexptected examples \citep{2011Sci...333.1602D, 2012Natur.481..475W}, important insights into planet formation have already been gleaned from the analysis \citep{2011arXiv1108.5842W}. 

\begin{figure}
\includegraphics[width=1\columnwidth]{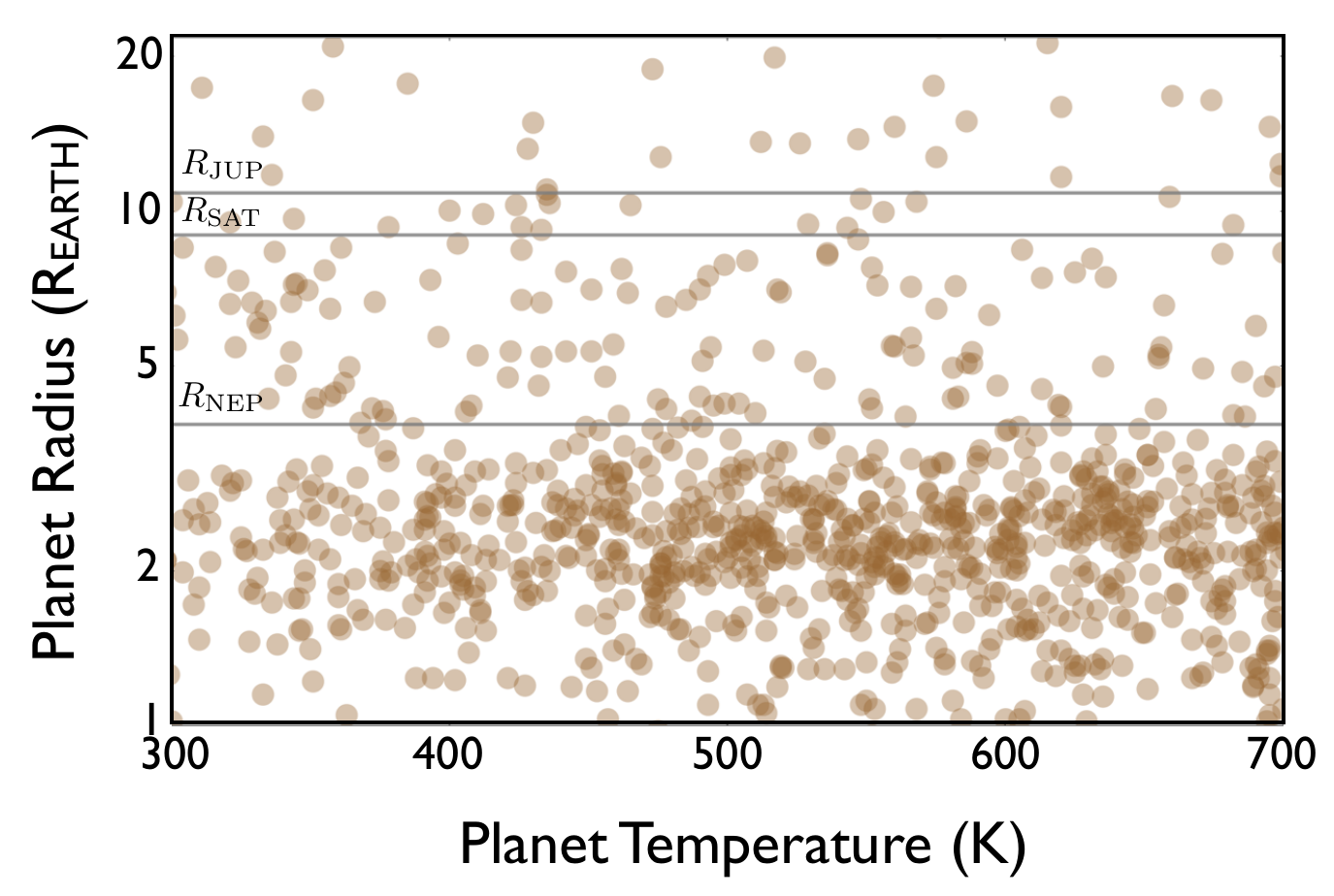}
\caption{Planetary radii as a function of planetary equilibrium irradiation temperature in the $Kepler$ sample. Note the considerable presence of giant-planet-like radii in this irradiation regime. The data was obtained from http://planetquest.jpl.nasa.gov/kepler.}
\label{keplerdata}
\end{figure}

It is interesting to note that the $Kepler$ data set contains objects whose radii are similar to that of Jupiter (and in a few cases even exceed it substantially), in the moderate irradiation range ($200$K $\lesssim T_{\rm{eq}}\lesssim800$K) (see Figure \ref{keplerdata}). Specifically, the latest application of the pipeline to the sample suggests that of 1333 total planetary candidates in this $T_{\rm{eq}}$ range, 68 have radii in the ($R_{\rm{SAT}}\lesssim R\lesssim 2R_{\rm{JUP}}$) range and 25 have radii that exceed $R_J$ by more than a factor of 2 \citep{2012arXiv1202.5852B}. Although the analysis of \citet{2011ApJS..197...12D} suggests that a dominant portion of the excessively large objects in the Kepler inventory are false positives, the physical nature of objects characterized by Jupiter-like radii is of considerable interest. Nevertheless, even basic information such as the average density is difficult to acquire since the overwhelming majority of the stars in the $Kepler$ field are rather faint, making radial-velocity follow up observationally expensive. Barring (near-)resonant systems, where transit timing variations can be significant \citep{2005Sci...307.1288H}, this means that the masses of the planets within the $Kepler$ sample will remain observationally unconstrained and theoretical inquiries are desirable. 

It is well known that giant planets comprising hundreds of Earth masses can have large radii, that exhibit only weak dependent on mass, and are instead primarily controlled by their chemical composition and the interior thermal state (Zapolsky \& Salpeter 1969; Stevenson 1982a). Furthermore, it is firmly established that radii of gaseous planets can increase with decreasing mass, thanks to the associated softening of the equation of state (Stevenson 1982a). Although, as illustrated by the wide-ranging numerical calculations of \citet{2007ApJ...668.1267F}, whether the radius increases or decreases with mass and the extent to which it does so, are rather sensitive to the amount of irradiation received by the planet as well as its chemical composition. 

In extreme proximity to the host star, the upturn in radius is well pronounced. For example, an evolved 20$M_{\oplus}$ planet irradiated at $T_{\rm{eq}}\simeq 1300$K is roughly twice as large as its isolated counterpart \citep{2008A&A...482..315B}. Depending on the planetary age, at even higher temperatures (e.g. $T_{\rm{eq}}=2000$K), the discrepancy may be as large as a factor of a few \citep{2005AREPS..33..493G}. On the other hand, planetary radii at $T_{\rm{eq}}\lesssim 100$K, do not differ from those of isolated objects much \citep{2007ApJ...668.1267F}. The current observational frontier lies in between these extremes, and to date, with the exception of only a handful of studies (e.g. \citet{2011ApJ...738...59R}), this parameter regime remains largely unexplored. In this study, we shall perform calculations that will place meaningful constraints on the mass-radius relationship of sub-Saturnian objects in the moderate irradiation regime. Specifically, \textit{the identification of a physically sound upper limit to the planetary radii at low masses and moderate equilibrium temperatures is the primary aim of this study.}

The possible range of chemical compositions of planets is generally not well known. However, the relatively low densities exhibited by some members of the well-characterized subset of the \textit{Kepler} catalog suggest that low overall metallicities cannot be ruled out. A particularly important example is the planet \textit{Kepler}-30d \citep{2012Natur.487..449S} which as we show below, has an envelope whose density does not exceed that of a cosmic H/He mixture substantially and cannot possess a core as massive as that typically invoked in the core-accretion model of planet formation \citep{1996Icar..124...62P}. Thus, motivated by the inferred structure of \textit{Kepler}-30d, we shall limit ourselves to a consideration of the most favorable planetary compositions for the fabrication of large radii. That is, for definiteness and simplicity, here we focus on planets with well-defined cores and H/He gaseous envelopes, though it is possible that in real objects the core material is partially mixed into the envelope\footnote{Mixing of core and envelope may not change the radius much for Jupiter mass planets when they are compared at similar temperatures. However, the consequences of this mixing are in general not simple for the radius-mass relationship because it affects the cooling history of the planet as well as the density distribution for a given temperature.} \citep{2012A&A...540A..20L}.

The paper is organized as follows. In section 2, we describe the setup of our numerical experiments and  perform simulations of planetary thermal evolution to explore the interior structure of \textit{Kepler}-30d. In section 3, we extend our calculations to lower masses and construct generic mass radius relationships, constrained by the hydrodynamical stability of the considered planets. We conclude and discuss our results in section 4.

\section{The Structure of \textit{Kepler}-30d}

Following initial detection \citep{2012arXiv1202.5852B}, the \textit{Kepler}-30 system was studied in greater detail by \citet{2012Natur.487..449S}, who determined the mass, radius, equilibrium temperature and age of \textit{Kepler}-30d to be $M=23.1\pm2.7M_{\oplus}$, $R=8.8\pm0.5R_{\oplus}$, $T_{\rm{eq}}=364$K (assuming zero albedo) and $2.0\pm0.8$ Gyr respectively. Here, we shall adopt the observed best fit parameters at face value for the generation of interior models.

Naturally, any model we consider is subject to hydrostatic equilibrium. With the knowledge of the equation of state and an assumed radiative structure of the atmosphere, the construction of a static interior model is relatively straight forward. This is however not enough, since the thermal state of the planet changes in time due to radiative losses of the interior entropy \citep{1999Sci...296...72G}. By extension, the planetary radius also contracts. Thus, in order to obtain definitive results that are characteristic of multi-Gyr old planets, evolutionary calculations of planetary structure are required. 

For our numerical experiments, we utilized the MESA stellar and planetary evolution software package \citep{2011ApJS..192....3P,2013arXiv1301.0319P}. Following \citet{2001ApJ...548..466B}, all of our models comprised constant density ($\rho_{\rm{core}}=5$ g/cc) solid cores embedded in gaseous H/He envelopes. The baseline heat-flux arising from radioactive decay within the cores was taken to be $10^{-7}$ ergs/s/g, similar to that of the Earth. The envelope metallicity was varied between $Z=0.02$ and $Z=0.04$, while solar $Y=0.27$ and slightly super-solar $Y=0.35$ values of the He mass fraction were explored. We note that if hydrodynamic mass-loss played a significant role in shaping the planetary structure \citep{2013arXiv1303.3899O}, a super-solar value of $Y$ can in principle originate from a preferential blow-off of Hydrogen.

The analytical radiative equilibrium model of the outer atmosphere was adopted from the work of \citep{2011A&A...527A..20G}. In the radiative portion of the atmosphere, following \citet{2010A&A...520A..27G} we take the constant visible and infrared opacities to be $\kappa_{\rm{V}}=10^{-2}$ cm$^{2}$ g$^{-1}$ and $\kappa_{\rm{IR}}=4\times10^{-3}$ cm$^{2}$ g$^{-1}$ respectively. These choices yield the closest agreement between the analytical radiative model used here and the state of the art numerical models of \citet{2008ApJ...678.1419F}. At optical depths much greater than unity, the tabulated Rosseland mean opacities of \citet{2008ApJS..174..504F} were used and the radiative convective boundary was computed as dictated by the Schwartzchild criterion. The reported radius of a given planet was taken to be the value corresponding to a chord optical depth of unity in visible light. 

\begin{figure}
\includegraphics[width=1\columnwidth]{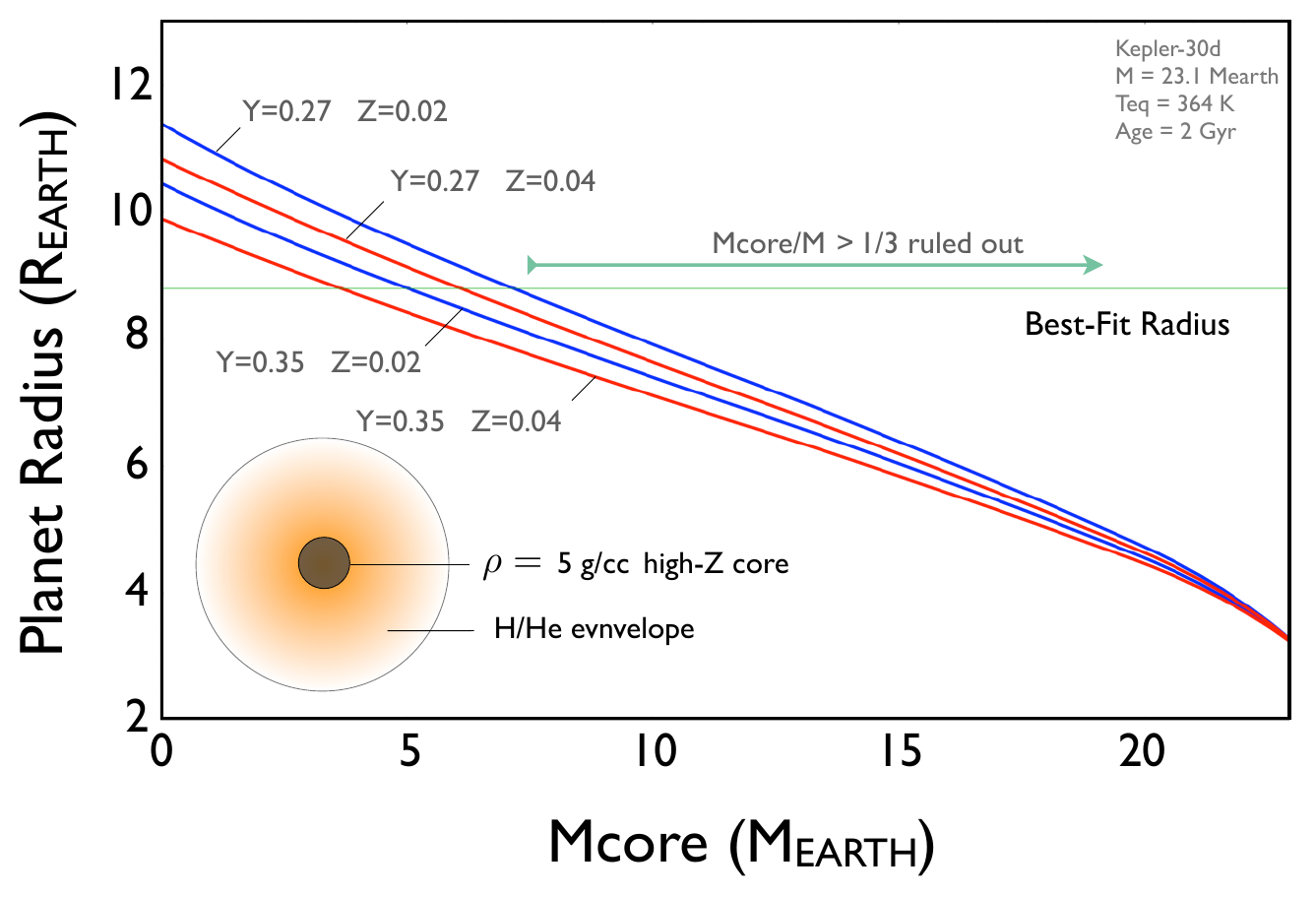}
\caption{The radius-core mass relationship for \textit{Kepler}-30d, assuming various envelope compositions. Blue curves correspond to envelopes with $Z=0.02$ while red curves correspond to $Z=0.04$. The cartoon in the left-bottom corner of the figure represents the considered two-layer interior models (here drawn to scale with a $5M_{\oplus}$ core).}
\label{kepler30}
\end{figure}

The search for admissible models of \textit{Kepler}-30d was performed in the following way. For a given choice of $Y$ and $Z$, the core-mass was varied between $M_{\rm{core}}=1M_{\oplus}$ and $M_{\rm{core}}=23M_{\oplus}$. The resulting initial conditions were integrated forward in time, yielding a sequence of model radii that decrease monotonically with $M_{\rm{core}}$. Importantly, radii also decrease monotonically with enhanced mean molecular weight, which means that there exists a maximum value of $M_{\rm{core}}$ above which the planetary radius cannot be matched. There also exists a maximal extent to which the mean molecular weight of the envelope can exceed that of a cosmic H/He mixture. However, such core-less solutions are strongly disfavored because the mass of $Kepler$-30d is too low for formation by gravitational instability to be plausible.

The $R-M_{\rm{core}}$ sequences for various compositions are shown in Figure (\ref{kepler30}). Adopting a solar composition envelope yields an upper bound on the core-mass of $M_{\rm{core}}\simeq7M_{\oplus}$. Meanwhile, the corresponding value for a $Y=0.35,Z=0.04$ envelope is a mere $M_{\rm{core}}\simeq3.5M_{\oplus}$. Unfortunately, because the planet's gravitational harmonics are not known, no useful lower bound on $M_{\rm{core}}$ can be formulated. It is noteworthy however, that the upper bound on $M_{\rm{core}}$ is surprisingly low.

The dominantly gaseous interior structure we obtained for \textit{Kepler}-30d is in sharp contrast with the inferred heavy element-dominated interior structures of Neptune and Uranus \citep{2011ApJ...729...32F}. This suggests that the diversity in composition and overall interior structure of low-mass planets is generally much more extensive than what is captured within modern state-of-the-art core accretion models. More specifically, this implies that the nucleated instability mechanism can operate even for comparatively small cores.

\section{Generic Mass-Radius Relationships}

Motivated by the results attained above, in this section we construct generic mass-radius relationships for evolved planets with a specified composition, extending down to minimum feasible masses. The radius-mass sequences were generated in a similar manner to the numerical experiments reported in the previous section. However, with the aim to constrain the planetary radii from above, the compositions of the envelopes were kept solar ($X = 0.71,Y=0.27,Z=0.02$) across the models. The core mass was varied between $1,3$ and $5\ M_{\oplus}$, while the total planetary mass range of up to 0.1 $M_{\rm{JUP}}$ ($\sim33M_{\oplus}$) was explored. 

Conventional generation of initial conditions within the framework of thermal evolution calculations is known to encounter numerical instabilities at sufficiently low masses. Consequently, here the initial conditions were constructed by imposing a slow mass-loss on a $M=0.1M_{\rm{JUP}}$ model. After the desired mass was attained, we imposed energy dissipation to the core and re-heated the gaseous envelope to the point where the thermal and gravitational energies of the body are comparable. The duration of the evolutionary sequences was formally taken to be $5$ Gyr. However, it should be noted that the changes in planetary structure were relatively small after the first $\sim$Gyr of integration. Likewise, we found the evolved radii to be largely independent of the detailed state of the initial condition, in agreement with published literature \citep{2001ApJ...548..466B, 2005Icar..179..415H}. 

Not all generated planetary models are guaranteed to be long-term stable. Indeed, some of the models we constructed were characterized by radii, exceeding that of Jupiter by as much as a factor of a few, rendering their stability against evaporation questionable. Accordingly, we formulated a criterion for model rejection in terms of the mass-loss rate due to atmospheric escape. 

Irradiated extrasolar planets can be susceptible to mass-loss due hydrodynamic winds originating in the upper atmosphere \citep{2009ApJ...693...23M, 2010A&A...516A..20V, 2012ApJ...761...59L}. Such winds are generated through the photoionization of H (and the associated heating) by extreme ultraviolet radiation. Provided that downward conductive heatflux or radiative cooling by H$_3^{+}$ is not overwhelming \citep{2009ApJ...693...23M}, the characteristic timescale for energy-limited evaporation is given to an order of magnitude by \citep{1981Icar...48..150W, 2008SSRv..139..437Y}
\begin{equation}
\tau_{\rm{e-lim}} \sim \frac{G M^2 K_{\rm{tide}}}{ \epsilon \pi F_{\rm{EUV}} R_{\rm{EUV}}^3},
\end{equation}
where $G$ is the gravitational constant, $F_{\rm{EUV}}=4.1(a/\rm{1AU})^{-2}$ erg s$^{-1}$ cm$^{-2}$ is the typical extreme ultraviolet flux of a $5$ Gyr old Sun-like star \citep{2005ApJ...622..680R, 2010A&A...511L...8S}, $K_{\rm{tide}}=1-3(R_{\rm{EUV}}/R_{\rm{Hill}})/2+(R_{\rm{EUV}}/R_{\rm{Hill}})^3/2$ is a geometrical factor that accounts for the fact any given parcel of gas only needs to reach the Hill radius to escape \citep{2007A&A...472..329E}, and $R_{\rm{EUV}}$ is a radius at which the atmosphere becomes optically thick to extreme ultraviolet radiation i.e. $n \mathcal{H}\sigma_{\rm{XUV}}\sim1$ where $n$ is the atmospheric number density, $\mathcal{H}$ is the scale-height, and $\sigma_{\rm{XUV}}\simeq10^{-18}$ cm$^2$ is the photoionization cross-section for Hydrogen \citep{2009ApJ...693...23M}. Meanwhile, $\epsilon\simeq0.25$ is a factor that parameterizes the efficiency of atmospheric escape. 

The EUV flux is considered to be constant here since we are not seeking to model loss and stability during early epochs of evolution. However, any model that we deem stable at $t=5$ Gyr will also likely be stable at any time greatly exceeding the T-Tauri phase of the evolutionary sequence (e.g. $t\gtrsim100$ Myr), because we generally find characteristic loss timescales of order $\tau_{\rm{e-lim}}\sim100$ Gyr or greater.

In all our models, $R_{\rm{EUV}}$ never exceeded the exobase (a radius at which the molecular mean free path becomes comparable to $\mathcal{H}$), meaning that the atmospheres were never truncated by Jean's escape. However, for certain models, $R_{\rm{EUV}}$ exceeded $R_{\rm{Hill}}$, implying mass-loss by Roche-lobe overflow. In such cases the characteristic evaporation timescale is given by \citep{1975ApJ...198..383L, 2010ApJ...721..923L}
\begin{equation}
\tau_{\rm{Roche}}\sim\frac{G M^2}{\pi\rho_{\rm{R}_{\rm{Hill}}}c^2a^3},
\end{equation}
where $\rho_{\rm{R}_{\rm{Hill}}}$ is the atmospheric density at the Hill radius and $c$ is the speed of sound. Generally, mass-loss by Roche-lobe overflow is orders of magnitude faster than that by extreme ultraviolet radiation-driven winds. Although any criterion based on the above estimates is only accurate to within an order of magnitude or so, we find this to be sufficient for our purposes, as we typically find a rapid transition from $\tau\gg$ Gyr to $\tau\ll$ Gyr across two models that neighbor each-other in mass. 

\begin{figure}
\includegraphics[width=1\columnwidth]{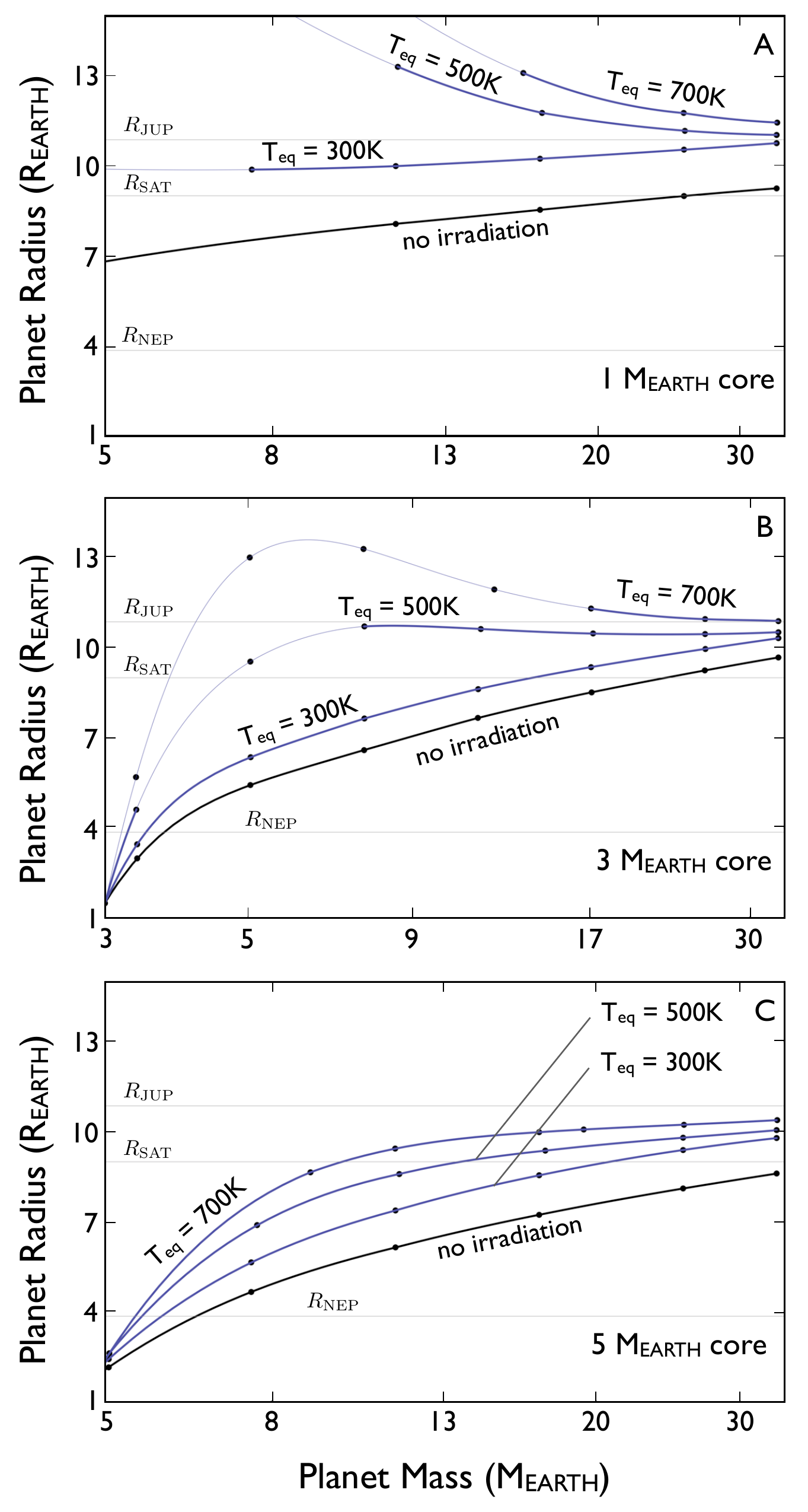}
\caption{Mass-Radius relationships of low-mass, gas-dominated planetary models. Panels A, B, and C correspond to planets with core-masses of $M_{\rm{core}}=1,3$ and $5M_{\oplus}$ respectively. On each panel, mass-radius relationships corresponding to equilibrium irradiation temperatures of $T_{\rm{eq}}=300,500$ and $700$K are shown as blue lines. Additionally, isolated mass-radius relationships are shown as black lines. Solid lines run through models that are stable against evaporation while the converse is true for thin lines. Note that radii characteristic of giant planets are readily attainable for mildly irradiated  $M\sim10M_{\oplus}$, $M_{\rm{core}}=1,3M_{\oplus}$ planets.}
\label{massradius}
\end{figure}

The mass-radius relationships for planets with core masses of $1,3$ and $5 M_{\oplus}$ are presented in panels A, B, and C of Figure (\ref{massradius}) respectively. Black dots represent the radii obtained through numerical experiments while the curves depict interpolation functions that run through the data. Thick curves imply models that are secure against evaporation while thin lines depict unstable models. In addition to the irradiated models (shown with blue lines), isolated (i.e. no irradiation) models are also presented  and are shown with black lines. For reference, Jupiter's, Saturn's and Neptune's radii are also marked.

The results highlight the fact that accounting for stellar irradiation, giant planetary radii can persist to surprisingly low masses (that is, $M\lesssim10M_{\oplus}$). Figure (\ref{massradius}) further affirms that the behavior of planetary structure is largely dictated by the associated core mass. Note that all models with a $5 M_{\oplus}$ core are stable against evaporation and roughly follow the cold (i.e. isolated) mass-radius relationship. On the contrary, $1M_{\oplus}$ core models are largely unstable below $M \lesssim15M_{\oplus}$ but can have radii comparable to that of Jupiter prior to the onset of evaporation. A similar scenario is observed for the $3M_{\oplus}$, $T_{\rm{eq}}=500$K set of models. Indeed, these models are essentially always characterized by $R\simeq R_{\rm{JUP}}$ above $M\gtrsim8M_{\oplus}$.

It is interesting to note that some of our models (e.g. those corresponding to 1 and $3M_{\oplus}$ and $T_{\rm{eq}} = 700$K) have radii that are bigger than that of Jupiter. As already discussed above, this upturn in radii is a direct consequence of the softening of the equation of state at lower pressures (an ideal gas has a softer equation of state than the deep interior of Jupiter). While reminiscent of the inflated Hot Jupiter radii \citep{2005AREPS..33..493G,2010SSRv..152..423F}, these objects are fundamentally different, since they require no additional heat sources or mechanisms for stalling gravitational contraction. That said, it is unclear if such objects are particularly significant within the context of the observational sample, since the models that show such an excess are close to the evaporation boundaries of the mass-radius diagrams. In fact, accounting for coupled evolution of gravitational contraction and mass-loss in a more self-consistent matter will likely yield an exclusion region that is a bit larger than what is shown in Figure 3. 

\section{Discussion}

In this letter, we have examined the structure of moderately irradiated low-mass low-density extrasolar planets. We began by analyzing the interior of a comparatively well characterized planet \textit{Kepler}-30d, and showed that the planet is likely composed of an extensive gaseous H/He envelope, surrounding a core that makes up less than a third of its total mass. Although qualitatively this object resembles a scaled down version of Saturn, it is important to recall that the mass of \textit{Kepler}-30d is typical of much more metal-rich objects such as Uranus or Neptune. The existence of \textit{Kepler}-30d immediately suggests that range of planetary interior configurations that occur in nature is much wider than that available for detailed study within the realm of the solar system.

Prompted by this notion, we extended our calculations to quantify planetary mass-radius relationships for cored low-mass gaseous planetary objects of solar composition. Our calculations underline the importance of stellar irradiation on the evolutionary tracks of low-mass objects. In particular, the constructed mass-radius relationships suggest that the radius of an irradiated body may exceed that of its isolated counterpart by as much as a factor of $\sim 2$ (e.g. the case of  $M\simeq10M_{\oplus}$, $M_{\rm{core}}=3M_{\oplus}$, $T_{\rm{eq}}=500$), bringing the radius well into the characteristic giant planet range. Collectively, our results suggest that extreme care must be taken in the interpretation of giant transit radii from the $Kepler$ sample, since the mass range corresponding to such radii can be quite extensive (i.e. spanning almost two orders of magnitude). 

One may wish to argue against a significant population of bodies like those considered in this work based on the (im)probability of their formation, since the gaseous component of our models is much enhanced over the standard models of typical objects in the considered mass range. Indeed, it is often said that one must have a "critical" core mass of $M_{\rm{core}}\gtrsim10M_{\oplus}$ in order to trigger gas accretion. However, this claim is ill-founded and is not actually relevant since a hydrostatically supported atmosphere around a core can be more massive than that envisioned within the context of the standard models \citep{1996Icar..124...62P} if either accretion is slower, the molecular weight of the envelope is larger, or the opacity is increased. This is evident for example in the simple analytical models of Stevenson (1982b) (see also \citet{2006ApJ...648..696I, 2009Icar..204...15B}). Protoplanets may also have circumplanetary disks that qualitatively change the characteristic accretion pattern and affect the planetary energy loss. Furthermore, alternative formation scenarios could likely be envisioned, a speculative example being one where objects of this type are sculpted out of more massive planets by intense ultraviolet-driven mass-loss during the first $\sim100$Myr of the stellar lifetime. Indeed, such scenarios have already been proposed in the exo-planetary context \citep{2006A&A...450.1221B, 2012ApJ...761...59L}.

Ultimately, our aim here is not to argue for or against any particular formation scenario for sub-Neptune mass gas-dominated planets. Rather, similarly to what has been done for \textit{Kepler}-30d, we propose that their existence can be validated or ruled out observationally. Beyond standard methods like transit timing variations, the most obvious approach to this is through radial-velocity monitoring of transiting planets. That is, if a giant-plenet-like radius is firmly established for a given object through transit observations but a commensurate radial velocity signal is not observed in the host star, such an object is likely characterized by a very low mass. Another approach to mass discrimination is exclusively photometric, and takes advantage of dependence of the transit radius on spectral frequency. Although the atmospheric scale-heights of Hot Jupiters comprise $\sim1\%$ of their radii at most, for $\sim10M_{\oplus}$ (albeit a factor of $\sim3$ cooler) planets with similar radii, the scale height is increased by about an order of magnitude. As a result, the chord optical depth of unity may correspond to substantially different radii in visible and infrared light. Both of these observational avenues should become readily available as the radial velocity precision continues to improve and future space-based missions such as JWST commence. \\
\\

\textbf{Acknowledgments} We thank Tristan Guillot, Geoff Blake, Ruth Murray-Clay, Adam Burrows and David Kipping for numerous useful conversations. We are grateful to the referee for a careful and insightful report that has greatly increased the quality of the manuscript. K.B. acknowledges the generous support from the ITC Prize Postdoctoral Fellowship at the Institute for Theory and Computation, Harvard-Smithsonian Center for Astrophysics.

\end{document}